\documentclass[sigconf]{acmart}
\renewcommand\footnotetextcopyrightpermission[1]{} 
\AtBeginDocument{%
  \providecommand\BibTeX{{%
    \normalfont B\kern-0.5em{\scshape i\kern-0.25em b}\kern-0.8em\TeX}}}

\acmConference[-]{-}{-}{-}

\begin{document}

\title{A Grounded Theory of Cognitive Load Drivers in Novice Agile Software Development Teams}


\author{Daniel Helgesson}
\affiliation{%
\institution{Dept. of Computer Science, Lund University}
\city{Lund}
\country{Sweden}}
\email{daniel.helgesson@cs.lth.se}

\author{Daniel Appelquist}
\affiliation{%
\institution{Softhouse AB}
\city{Malm{\"o}}
\country{Sweden}}
\email{daniel.appelquist@softhouse.se}

\author{Per Runeson}
\affiliation{%
\institution{Dept. of Computer Science, Lund University}
\city{Lund}
\country{Sweden}}
\email{per.runeson@cs.lth.se}


\begin{abstract}
\textbf{Context:} Agile software development in teams is a socio-technical iterative activity, involving a network of people, software and hardware. The capacity of the human working memory is limited, and cognitive load is induced on developers by tasks, software systems and tools.

\textbf{Objective:} The purpose of this paper is to identify the largest cognitive challenges faced by novices developing software in teams.

\textbf{Method:} Using Grounded Theory, we conducted an ethnographic study for two months following four ten-person novice teams, consisting of computer science students, developing software systems.

\textbf{Result:} This paper identifies version control and merge operations as the largest challenge faced by the novices. The literature studies reveal that little research appears to have been carried out in the area of version control from a user perspective.

\textbf{Limitations:} A qualitative study on students is not applicable in all contexts, but the result is credible and grounded in data and substantiated by extant literature.

\textbf{Conclusion:} We conclude that our findings motivate further research on cognitive perspectives to guide improvement of software engineering and its tools.

\end{abstract}



\maketitle

\section{Introduction}
With the advent of agile methodologies, the socio-technical character of software engineering has been further promoted. With its emphasis on ``individuals and interactions \emph{over} processes and tools''\footnote{\url{https://agilemanifesto.org}}, the Agile manifesto makes a strong promotion of the social side. The technical side of the phenomena seems to be more studied than the social side~\cite{lenberg_behavioral_2015}~\cite{storey_who_2020}, although human factors were part of software engineering already from its inception. For example, \emph{personnel factors} were discussed at the NATO conference 1968~\cite{Nato1968}. In their summary of 50 years of the psychology of programming, Blackwell et al. synthesize research in the intersection between several research communities~\cite{Blackwell2019}. Referring to B{\o}dker~\cite{Bodker2015}, they see three major waves of research, namely ``\emph{in the first wave, cognitive psychology and human factors; in the second wave, social interaction within work settings; and in the third wave, a focus on everyday life and culture.}''

As a contrast to the third wave, which is scaling down towards the craft or hobby programming, we would like to bring the up-scaling of software engineering into attention. The complexity added by composing systems of systems, the increasing and constant change enabled by agile methodologies and DevOps, the more and more powerful tools for software engineering -- all add to the cognitive load of developers and managers of software engineering projects. Blackwell et al. claim in their proposal for a future research agenda, that \emph{```[d]esign thinking', solving `wicked' problems, and reasoning more broadly about software systems and systems-of-systems has not received the sort of attention that has been devoted to, say, program comprehension.}''~\cite{Blackwell2019}

In an attempt to more broadly adress the cognitive load induced from these trends, Helgesson et al. studied \emph{cognitive load drivers} in large scale software development~\cite{helgesson_cogload} and found three clusters of drivers, namely \emph{tools}, \emph{information}, and \emph{work \& process}. The authors also noticed that the temporal perspective of software development, particularly revision control, created specific problems. 

To further advance the understanding of cognitive load in software engineering, we set out to ethnographically study agile software engineering projects, using grounded theory. As we hypothesise that some of the cognitive leads are compensated and mitigated through increased experience and work-arounds learned over the years, we choose to study novice software engineers~\cite{students_exp}. Our study context is quite advanced for novices, an agile software engineering course, running for 14 weeks, in which students work in 10 person teams in a simulated work environment, adhering to XP principles. We observe four teams out of a total of twelve teams participating in the course. 

Our research goals are to identify the most dominant \emph{cognitive load drivers} and to observe differences and similarities between groups with different characteristics. We combine the teacher role of on-site customer with the ethnographer role, taking field notes of the observations. Further, we collect weekly questionnaires, short reflection notes from the students, and arrange a focus group discussion with each team. We use grounded theory practices in coding all the material, from which our theory emerges. 

We conclude that version control, branching and merge operations is the dominant load factor in the projects, and thus we explore this phenomenon in detail. 

\section{Method}
\label{sec:Method}
\subsection{Grounded theory}
Grounded theory (GT) is a systematic and rigorous methodological approach for inductively generating theory from data~\cite{glaser_strauss_discovery}~\cite{charmaz_constructing}~\cite{stol_gt}. Stemming from social sicences, GT was developed by sociologists Glaser and Strauss, as a qualitative inductive reaction to the quantitative hypotetico-deductive research paradigms dominant in the 1960s. The main difference, apart from being qualititative rather than quantitative, is that the purpose of GT aims at generating theory, rather than to be used as an instrument for validation or testing of theory~\cite{stol_gt}. It is iterative and explorative~\cite{charmaz_constructing} in nature, and thus suitable for answering open ended questions such as \emph{what's going on here?}~\cite{stol_gt}\cite{adolph_gt}.  
Specific guidelines and suggestions for GT research in Software Engineering are provided by Stol et al.~\cite{stol_gt}.

We primarily opted for Charmaz GT handbook~\cite{charmaz_constructing} as guidelines, complemented by earlier works by Glaser~\cite{glaser_78}~\cite{glaser_92}. Specifically we used grounded theory ethnography~\cite{charmaz_mitchell} -- an approach that gives  \emph{``priority to the studied phenomenon or process -- rather than the setting itself}''~\cite{charmaz_constructing}. The ethnographic approach allows for exploring not only \emph{what} practitioners do, but also \emph{why} they do it~\cite{sharp_ethno}. Core elements in the ethnographic approach is the empathic approach \emph{to describe another culture from the members' point of view} and the intrinsic \emph{analytical stance}~\cite{sharp_ethno}. As with grounded theory, modern ethnography also stems from  social sciences~\cite{sharp_ethno}. Not extensively used in Software Engineering~\cite{sharp_ethno}, it has however been used to study agile teams~\cite{sharp_distributed_1}\cite{sharp_ethnographic_xp}. 

\subsection{Research goals}
Central to `original' Glaserian GT and Charmaz Constructivist GT is that the actual/final research questions are not defined up front. Glaser suggests that the researcher should start with an \emph{area of interest}~\cite{glaser_92}~\cite{stol_gt}, while Charmaz suggestion is that the researcher should start with \emph{initial research questions} that \emph{evolve} through the study 
~\cite{charmaz_constructing}~\cite{stol_gt}. We decided to pursue two open ended research goals:

\begin{itemize}
\item[A)] To identify the most dominant \emph{cognitive load drivers} from the \emph{novice point of view}, and
\item[B)] To observe \emph{differences} or \emph{similarities} between different \emph{group compositions}.
\end{itemize}

\subsection{Case description}
\label{sec:case}
The course that we used as study object is a mandatory course for sophomore computer science\footnote{Translations of educations are difficult, in international terminology 'Computer Science' is as close as we can translate it. It is a five year master (engineering) program mostly aimed at software rather than hardware. The program resides at the Faculty of Engineering, and the program responsible reside at the department of Computer Science.} students aiming at teaching practical software development in teams using agile methodology, presented in detail by Hedin et al. \cite{hedin_1}. The course runs for two terms (14 weeks) and consists of one study block (seven weeks) consisting of lectures and practical lab work, and one study block (seven weeks) in which the students work together as 10 person teams, largely adhering to XP principles~\cite{sharp_ethnographic_xp} developing a software product. All teams develop a software system based on the same basic stories, but the stories are somewhat open ended leaving room for differentiation. The teams are coached by two senior students undertaking a course in practical software coaching, that runs in parallel for the same duration. PhD students serve as \emph{customer} for 3--4 teams each.

The teams develop their system for a term (seven weeks) in 6 full day sprints, each preceded by a two hour planning session in which the \emph{cost}/\emph{effort} for the user stories are estimated by the students and prioritised by the \emph{customer}. The students make 3--4 incremental releases during the project, roughly with a cadence of one release every two sprints. 

\subsection{Design considerations}
We opted for a flexible case study design~\cite{runeson_case}, to allow for improvisation based on observations and forces outside of our control (which once you take research into the wild are plentiful). Once in the field, flexibility becomes utterly important~\cite{sharp_ethno} as the researcher must be ready to adapt to changing situations quickly.

We had a strict time box for our field study, since the course executed over the duration seven weeks with one day sprints on Mondays, following a two hour planning session on the Wednesday before. Apart from the fixed schedule for observations we also had to take into account the work load of the students when injecting experiments and eliciting interviews. We had the ambition to cause as little disturbance as possible. In order to achieve triangulation we opted to collect as many data sources as possible.

We also decided to use \emph{distributed cognition}~\cite{hollan_dc} as initial lens, or filter, for our observations. Distributed cognition, further described in Section~\ref{sec:literaturereview}, is a branch of cognition studying cognitive processes distributed in groups rather than cognition from the individual perspective. While the use of an initial lens could be thought of by some readers as contradictory to the central tenet in grounded theory, we hold this (potential) critique as moot. We were targeting observations of cognitive load drivers in interconnected network of people and digital tools, so we needed some starting point for our observations.

\subsection{Group composition}
\label{sec:studentselection}
Firstly we anonymously picked 14 student candidates, based on a high grade (grade average in excess of 4.5 on five grade scale, where \emph{pass} is denoted as 3) in the first two programming courses, and a lower grade (i.e. \emph{pass} or \emph{incomplete}/\emph{fail}) grade in multidimensional calculus. Secondly we anonymously picked 14 student candidates based on a high grade (grade higher than \emph{pass} on five grade scale, where \emph{pass} is denoted as 3) in multidimensional calculus, regardless of their grade in programming courses.

The two anonymous candidate lists were then sent to the course responsible who then created one purposely selected group each out of the two candidate lists and two randomly selected groups. After this process we had four groups in total, i.e. four units of analysis within the case. It should be noted that the authors at no point in time were informed of what group consisted of what selection.

\subsection{Consent}
Together the course responsible and the first author ultimately reached the conclusion that the optimal solution (in regards to time constraints and complexity) was to inform the students in the four groups at the start of the course that we would be carrying out research throughout the course, describe the overall purpose/general research goal of the study, that we were looking at the groups and not the individual members and offer any student not willing to participate to change groups prior to the first sprint. No student asked to exchange groups.

In every interaction that was recorded or photographed, we actively asked every student participating for permission, while pointing out that everything expressed in the exchange would be anonymous and confidential, and that no recordings would be distributed outside of the three researchers participating in the study. For further ethical considerations, see Section~\ref{sec:ethics}.

\vspace{-3mm}
\subsection{Data collection}
The first and the second author followed all planning sessions in parallel. As we had to monitor sessions in parallel we opted to alternate between observing in pairs and by ourselves. All in all we covered 24 planning sessions where the first author actively participated in the meetings acting as \emph{customer on site} providing students with clarifications of stories, priorities etc, while the second author passively observed. After each session we spent, roughly, 15 minutes discussing what we had observed. Field notes were written by hand, and after the termination of the field work compressed in \emph{memo} form. 
The first author actively participated in all full day sprints while acting as \emph{customer on site}. The four teams were situated in two computer labs, allowing for observation of two teams simultaneously. Field notes were written by hand, and after the termination of the field work compressed in \emph{memo} form. We specifically opted to not be part of breaks, lunch hour etc. for respect of the students privacy. Since our research focus is the phenomenon of cognitive load from a team perspective, rather than team work in general, we do not see this as a threat to our observations. 

In addition, we added a weekly questionnaire to be filled out by each student after every sprint (all in all 4*10*6 questionnaires) in order to follow up on what we had observed so far throughout the project. The first two weeks the questionnaire targeted sources of information and information tools used by the students. In the third and fourth questionnaire we introduced check boxes and free text space, allowing the students to express what they perceives as the major problems they had been challenged by throughout the project. In the fifth questionnaires questions were added to capture the outcome of one of the experiments, see subsection~\ref{sec:experiment_testing}. The final questionnaire was extended with questions regarding team spirit and overall satisfaction. The aggregated response rate for all 24 sets of questionnaires (6 for each team) were 93\% (out of the 240  questionnaires we handed out we got 223 in return, and no single set had a lower response rate than 8/10). 

Further, as a requirement of the course all students wrote short individual reflections after each sprint, as a retrospect exercise. After the course we aggregated these pages, anonymised all content and created one .csv file per team with the content broken down in line-by-line format for \emph{open coding}.

After the final sprint we held one-hour focus group discussion with each team. The discussions took place in two by two parallel sessions. Two instances were held by the first author, one by the second and third author collectively and one by the second author. In order to keep the different sessions coherent and comparable we followed a semistructured manuscript containing four themes we had selected as emerging concepts from our observations. We used pair-wise post-it discussions, followed by group discussions where each pair reflected on what they had come up with. The post-it stickers were collected, numbered and digitized. Each session was also recorded using video and sound.

\subsection{Quasi-experiments}
Inspired the reasoning on \emph{ethnographically natural} field experiments by Hollan et al.~\cite{hollan_dc}, which corresponds to quasi-experiments in software engineering literature~\cite{wohlin_experiment_book_2012}, we decided to extend our data set with empirical data from three minor field experiments. These were dressed as improvised exercises, which is a part of the overall course concepts, where \emph{unplanned} customer changes could take place~\cite{hedin_1}.

\subsubsection{Quasi-experiment -- group constellation}
\label{sec:experiment_group}
Our first experiment consisted of creating four teams with different member compositions, with the purpose to see what differences, if any, we could observe during the observation study (and through the other data sources). See subsection \ref{sec:studentselection}.

\subsubsection{Quasi-experiment -- exploratory testing}
\label{sec:experiment_testing}
The second experiment consisted of assigning the students with a surprise story in preparation for the fifth sprint. The story consisted of little more than the instructions to: \emph{execute roughly 1 hour of exploratory user tests of the system under realistic race conditions using four team members documenting the issues encountered}, and further to reflect on the experience in their weekly reflections (that all students fill out after each sprint). The story was handed out during the planning session the week before the full day sprint during which it was planned. We collected information of the activity from questionnaires (Q5/Q6) and from discussions with students and coaches during the following sprint and planning session.

\subsubsection{Quasi-experiment -- merge-back}
\label{sec:experiment_merge}
The third experiment consisted of the request to implement two sets of changes, in two separate files, and upon completion of the first task request a merge-back and recreation of the first release. Each team was handed a story card describing the two code blocks to be implemented \emph{first thing in the morning} during the final planning session leading up to the final sprint. Each team was asked to notify their \emph{customer} upon completion of the task. In order not to compromise that functionality/integrity of their respective systems the two code blocks were dummy snippets that were commented out. The experiment was documented using video and sound recording.

\subsection{Analysis}
Given that we had a limited time window for our observations, we did not have a lot of time for analysis during the field work. We exchanged notes and discussed our observations over lunch breaks. After the field work was completed the first and second author started a more formal analysis stage.

\textit{Initial coding}~\cite{charmaz_constructing} -- the first and second author each performed open line-by-line coding of the student reflections and the post-it stickers. We then exchanged our reflections in short memo form. In parallel, the first author did an initial overview of the contents of the questionnaires.

\textit{Focused coding}~\cite{charmaz_constructing} -- the first and the second author had a two day session in which the questionnaires, focus groups post-it stickers and student reflections were analysed from multiple perspectives and the parts that we found relevant was extracted and documented digitally. We also extracted relevant `soundbites' from free text answers, and digitised them. The findings were condensed in a short memo.

\textit{Theoretical coding}~\cite{charmaz_constructing} -- the theoretical coding was executed by the first author, using Glaser's \emph{`6C'-coding family}~\cite{glaser_78} as a starting point, including conditions, causes, consequences, context, contingencies (or variables), and covariances. The work was done in memo form and visualized on an A1 sheet using postit stickers. After a few iterations of coding, sketching and memoing a theory was emerging. The first and third author had a one-hour session in which the theory was discussed from various angles and a few of the constructs were redefined. After this the first author did a minor rewrite of the theoretical coding memo.

\subsection{Theoretical saturation}
Having iterated through \emph{open coding} and \emph{focused coding} of the data set, we saw the need of further \emph{saturation}
in order to provide some more insight from \emph{the members' point of view}. In order to do so, we went through the recordings of the focus groups in order to provide some additional insight. Finally we reached out to a handful of students whom previously agreed to do minor follow up interviews. We held three short (15--20 minutes) open interviews specifically aimed at understanding what the students perceived as tool interaction related issues. The interviews were conducted by the first author and were documented by additional field-notes. All quotes and findings were reread to the subjects at the end of these interviews.

\subsection{Literature review}
In its original form, research questions in GT studies should emerge from the research, not be defined apriori~\cite{stol_gt} and \emph{extensive} literature should be avoided prior to the emerging of theory. That being said, Charmaz takes a more pragmatic stance on literature and research questions and emphasises the iterative nature GT, thus allowing for initial research questions that evolve through the research project as well as abductive reasoning on extant literature, recommending a \emph{preliminary} literature review \emph{``without letting it stifle your creativity or strangle your theory}''~\cite{charmaz_constructing}.

As a consequence we did an initial, rather limited, literature study of Distributed Cognition from a Software Engineering perspective. Following the coding cycles we did an additional, or final, literature review on the central phenomenon of the theory we generated, i.e. GIT, version control and  merge operations from a user perspective.
See Section \ref{sec:literaturereview} for findings.

\section{Analysis}
\label{sec:Analysis}
This section presents the theory generated from the dataset. Based on the findings from open and focused coding of our data set, the emerging concept we focused on was \emph{issues regarding version control, branching and merge operations}.

For the first attempt at formulating the theory, a theoretical conceptual explanation of what we observed, we based our theoretical coding on Glasers \emph{6 C-coding family}~\cite{glaser_78}~\cite{stol_gt}, while observing Thornberg and Charmaz reflection that the researcher should avoid being \emph{hypnotized} by Glaser's coding families~\cite{thornberg_charmaz_theoretical_gt}. This is analogous to Glasers argument that all codes should \emph{earn}~\cite{glaser_78} their way into the theory. Thus, we used \emph{the 6 C's} as a starting point, and allowed for modifications throughout the \emph{theoretical coding} phase.

\begin{figure}[t]
\begin{center}
\includegraphics[width=\columnwidth]{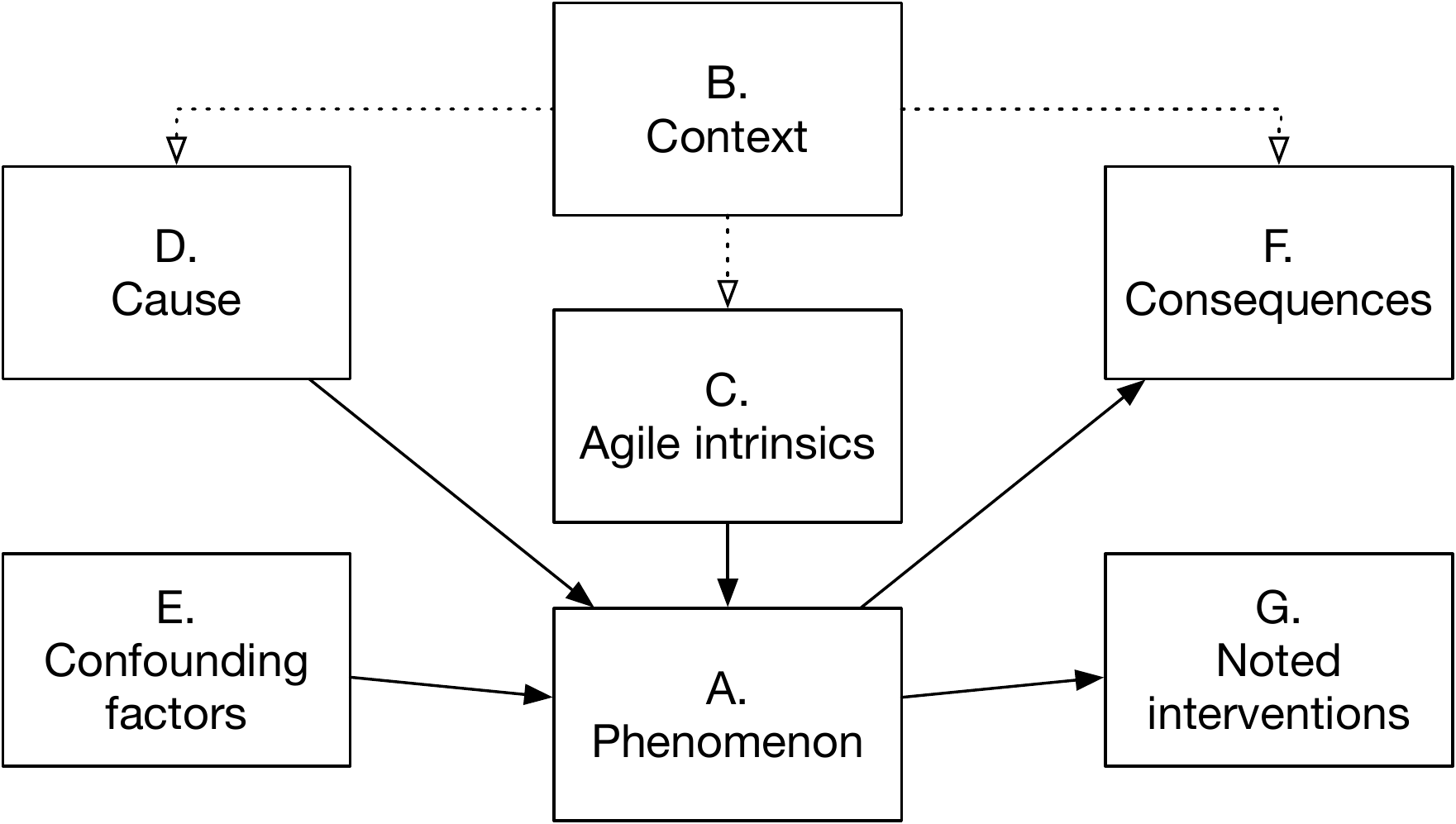}
\caption{Generated theory of the causal and consequential dimensions in regards to version control, branching and merge operations encountered in the projects.}
\label{fig:concepts}
\end{center}
\end{figure}

\begin{figure*}[t]
\begin{center}
\includegraphics[width=\textwidth]{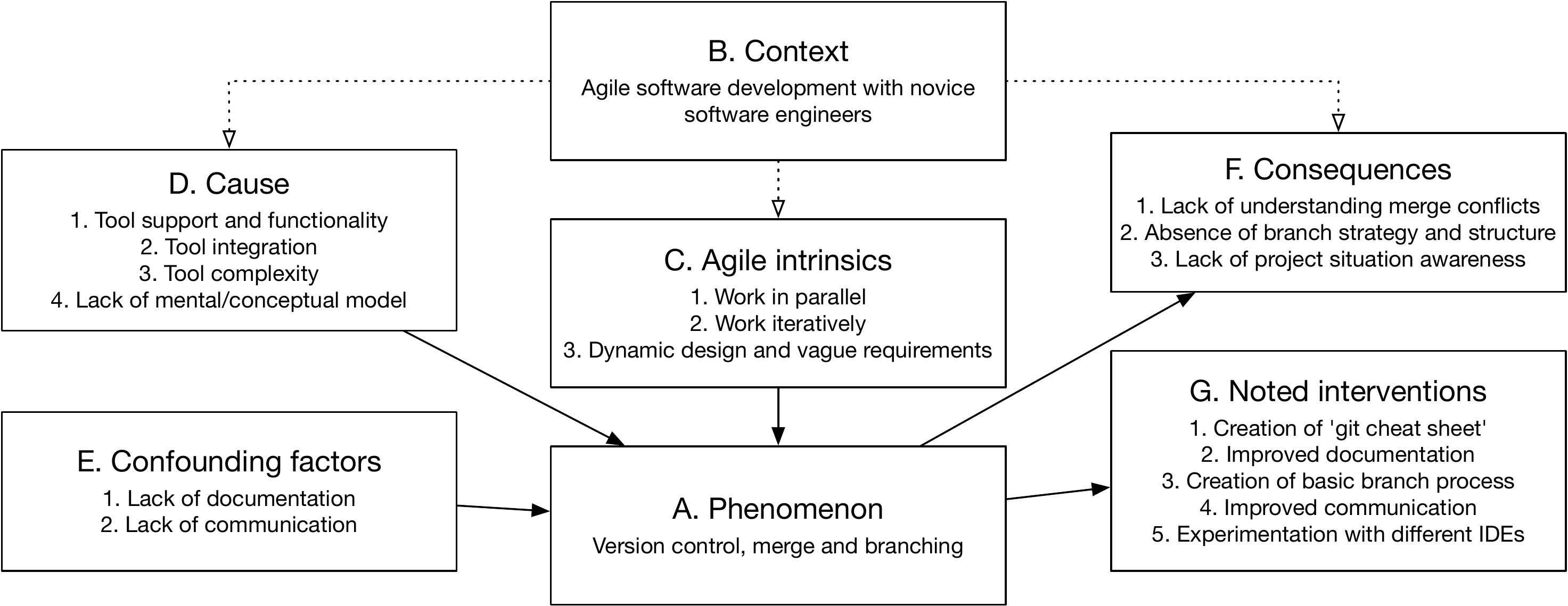}
\caption{Generated theory from Figure \ref{fig:concepts}, further extended with the detailed codes from the analysis.}
\label{fig:conceptsfull}
\end{center}
\end{figure*}

A conceptual rendering of our generated theory of \emph{issues regarding version control, branching and merge operations}, is illustrated in Figure~\ref{fig:concepts}. The center bottom rectangle (A) describes the core phenomenon, \emph{version control, branch \& merge issues}, while the other codes are represented by surrounding rectangles (B--G). Cause, correlation and effect are represented by arrows. Context is represented using dotted arrows. For each code a corresponding subsection is found below. Along with the analysis, the theory is detailed in Figure~\ref{fig:conceptsfull}.

Throughout the analysis section we provide examples of `quotes' from the data set. `S'/`I'  denotes interview subject and researcher respectively. We have added \emph{emphasis} for \emph{clarity} and occasional further clarifications within [hard brackets].

\subsection{Phenomenon (A)}
Throughout our observations (field notes) and our questionnaires we noted that version control, branching and merge operations caused a disproportionate amount of loss in productivity and time. The questionnaires for all teams systematically indicated \emph{version control}, \emph{branching} and \emph{merge conflicts} as the most disruptive challenges encountered throughout the project, and as a consequence this is the phenomenon we chose to explore. 

 E.g.: ``Git/Merge -- We are \emph{unsure} of how to \emph{use git properly}''  (from student questionnaires -- in response to what has been the biggest hurdle faced during the project).

\subsection{Context (B)}
We define the context from which our observations are extracted, and in which they are valid, as that of agile software development teams, consisting of novices, using Git. Admittedly, this could result in a rather narrow validity window in terms of generalization. However, in our experience (both the first and second author has 15+ years experience of professional tool driven software development in large/distributed software projects) this observation, practitioners struggling with Git is commonplace in industry. 
Further, using novices as study objects would rather reveal cognitive challenges, as these challenges are not mitigated by \emph{trained behavior}, \emph{learning effect} or \emph{status quo bias}. 

We created our data set from observing and interacting with four different teams of \emph{novice software developers} in parallel. All teams were using XP, and developed their system using the same basic stories/requirements (see Subsection~\ref{sec:case} for details). While tool chain set up and development environment (IDE) differed somewhat between the teams, all teams used Git hosted by Bitbucket for version control (albeit with different branch strategies).

In light of the observed lacunae in extant software engineering literature, we note that version control from a user perspective is an area not thoroughly studied in the research community. Those few studies we found systematically indicate that our observations are valid in a wider context.

\subsection{Agile intrinsics (C) -- \textbf{root cause} \& \textbf{driver}}
We identified the iterative and parallel nature of agile software development, which we refer to as \emph{Agile intrinsics}, as the underlying \emph{Root Cause} of the observed merge conflicts. For further clarity we added an additional subcode, \emph{Observed driver}, which refers to phenomena driving the cognitive load. In order to achieve better granularity we further break the \emph{agile intrinsics} construct up into three different subcodes: \emph{(Work) in parallel}, \emph{(Work) iteratively} and \emph{Dynamic design and vague requirements}, since they are related in terms of root cause but have quite different consequences. As indicated by the \emph{intrinsics} in the main category, these traits are inherent (largely by design) in the nature of agile software development. 

\subsubsection{(Work) in parallel} 

\textbf{Observed root cause:}
When starting up the project, the code base is very small, and different programming pairs are developing, and modifying the same code/classes/files, creating dependencies and diverging implementations ultimately leading to merge conflicts. While this to some extent was mitigated by adopting rudimentary branch strategies, the problem persisted throughout the projects.

We note that it appeared hard for the developers to find out \emph{who did what, when and why?}, ultimately leading to a lack of understanding of implementation details, or a micro perspective, thus making subsequent merge conflicts harder to resolve. We also noted that this caused the developers to implement their own variations of similar methods (e.g. utility methods).
E.g.: 
\begin{itemize}
\item[--] ``Trying to merge code that someone else has written.'' 
\end{itemize}(from student questionnaires and focus groups in response to what they found being difficult during their projects).

\textbf{Observed driver}: Diverging implementations leads to conflicting implementation details, further resulting in merge conflicts.

\subsubsection{(Work) iteratively}

\textbf{Observed root cause:}
The iterative nature of the development results in constantly shifting implementation details and this subsequently drives merge work. The constant change in code leads to a lack of understanding from a micro perspective, reimplementation and duplication of code as different development pairs reimplement existing functionality.
E.g.: 
\begin{itemize}
\item[--] ``Parts of code \emph{unknown}, having to interact with code that \emph{someone else has written}, better after refactoring.''
\end{itemize}(from student questionnaires and focus groups in response to what they found being difficult during their projects).

\textbf{Observed driver}: Refactoring implementation leads to changing implementation details, further resulting in merge conflicts.

\subsubsection{Dynamic design and vague requirements}
\textbf{Observed root cause:}
Since there is no set architectural design/framework, nor a complete set of requirement or user stories, in the beginning of the projects, there was no cohesive collective goal for the developers. Further, architectural changes drives extensive refactoring and results in subsequent merge conflicts. Despite the fact that this is an inherent feature of XP -- ``XP is a lightweight methodology for small-to-medium-sized teams developing software in the face of vague or rapidly changing requirements''~\cite{beck_1} -- it is none the less something we noted as a systematic cause of refactoring and merge conflicts.
E.g.: 
\begin{itemize}
\item[--] ``\emph{Hard} to change data structures. This causes merge conflicts and bugs. Improve communication [within the team]?''
\end{itemize}(from student questionnaires and focus groups in response to what they found being difficult during their projects).

\textbf{Observed driver}:  No set design at the beginning of projects leads to refactoring of structure and changing of architecture, resulting in merge conflicts.

\subsection{Observed cause (D) }
This part of the analysis provides a reasoning on our observations on the observed causes of merge incidents.
\subsubsection{The impact of tool support and tool functionality}
Throughout the study we noted that the students were quite opinionated about functionality and support of the different tools and how well they were integrated. All teams started their respective projects using Eclipse and GIT hosted on Bitbucket. Out of the four teams, two ultimately migrated from Eclipse to another IDE; intelliJ in one case and VSCode in one case. 


When discussing tools during focus groups the importance of the user support the developers experienced became obvious. We noted that ease of use, intuitive interaction and visual support and offloading was something the students noted as very important in terms of reducing cognitive load. 
This is illustrated below in an excerpt from a focus group dialogue between three students (SI--SIII) and the interviewee (I):

\begin{itemize}
\item[SI:] -- ``Many had problems \emph{seeing} what changes that were being made, that is when you fetch; it might be related to Eclipse [integration with git], it became better with VSCode, with the \emph{colours} [indicating visual offloading], the \emph{visual}, to be able to understand what has happened.''
\item[I:] -- ``But what \emph{experience} have you had in regards to the \emph{tool support} you have had in order to \emph{solve merge conflicts}?''
\item[SI:] -- ``Eclipse was really \emph{messy}.''
\item[SII:] -- 
``...it was really hard to see \emph{what} changes were coming from \emph{where} -- [extended pause, thinking] - and I think the \emph{colours} in VSCode are really good [indicating visual support]. You really \emph{see}, visually, \emph{what} is \emph{what}.''
\item[SIII:] -- 
``...when we were using Eclipse we \emph{switched to using the terminal} [use of GIT through command line interface (CLI) instead of IDE integration] instead, it just feels a lot \emph{easier}.''
\end{itemize}(Focus Group, excerpt from video recording T@12.43).

The importance of visual support became even more obvious during saturation:
\begin{itemize}
\item[S:] -- ``The merge support in VSCode is \emph{graphical} and easy to understand -- it is \emph{intuitive}.''
\item[S:] -- ``The \emph{merge support} in VSCode is very \emph{clear}, it provides help on resolving the conflict, it \emph{shows} \textless source code 1\textgreater~and \textless source code 2\textgreater~in the GUI and it is \emph{simple} to choose by clicking a button.''
\item[S:] -- ``intelliJ is actually, in my opinion, better than VSCode. It gives even better and \emph{more visual} merge support.''
\end{itemize}(Field notes -- saturation)

\subsubsection{Tool integration}
We decided to further break down the analysis of the tool support further, in order to be able differentiate different angles of the experience of the students. We noted that the actual integration of Git in the IDEs was considered quite important, and a contributing factor when it came to changing IDE.
\begin{itemize}
\item[S:] -- ``The \emph{graphical integration of Git} in Eclipse is \emph{difficult to understand.}''
\item[S:] -- ``Eclipse is \emph{complicated} in terms of Git integration, and it is \emph{easier to use} git through a \emph{terminal} than through Eclipse.''
\item[S:] -- ``The \emph{integration} between Git and VSCode is \emph{superior} to that of Eclipse.''
\end{itemize}(Field notes -- saturation).

\subsubsection{Tool complexity}
The actual importance of tool complexity came to some surprise to the first author. We observed several reflections on the intricacies and complexities of Git in the dataset. We found compelling evidence that the complexity of Git was indeed a main cause of concern and cognitive load for novices, but the intricacies of Git was not the only cause of concern -- the complexity of the IDE was also a definite issue and cause of confusion.
\begin{itemize}
\item[S:] -- ``Version Control -- Git is very \emph{difficult}.''
\item[S:] -- ``What would make Eclipse better? Better \emph{merge support} and better overview, making it \emph{easier to find functionality}.''
\item[S:] -- ``VSCode feels \emph{simpler}, with less functionality but it is a lot \emph{less overwhelming}. It has a lot better learning curve.''
\item[S:] -- ``Eclipse is \emph{complicated} and it is \emph{difficult to understand} the structure.''
\end{itemize}(Field notes -- saturation).

Somewhat counter intuitively we also observed the following reflections on Git the command line interface:
\begin{itemize}
\item[S:] -- ``Git/CLI [in terminal] is good because it looks the same in every environment.''
\item[S:] -- ``Git/CLI [terminal] is good because all git online resources describe Git through CLI, so it is a lot easier to copy a line of commands and paste it into the terminal than to try do do the same thing through a GUI.''
\end{itemize}(Field notes -- saturation).

\subsubsection{Lack of mental/conceptual model of version control and branch structure}
Based on the outcome of the merge experiment (subsection~\ref{sec:experiment_merge}), which we considered a trivial Git/branch operation, we noted that the students' understanding of reasonably straight forward branch operations in Git was somewhat limited. Out of the three groups that did the experiment (one team dropped out because of time constraints in their project), no one came up with a viable solution (albeit they came up with interesting and manually labour intensive ways to approach the task). At the end of the time-slot given for coming up with a solution, the first author provided a hint of the form ``Well, maybe you should google \emph{git squash} and \emph{git cherry pick}?''. Subsequently all three teams adequately solved the exercise in a matter of minutes.


\subsection{Confounding factors (E)}
\subsubsection{Lack of documentation}
We noted that a systematic lack of documentation (i.e. \emph{code comments}, \emph{commit messages}, \emph{design documentation}) plagued the groups throughout their respective projects. This added to the lack of understanding the merge conflicts. We also noted that the students became aware of these aspects and, to a varying degree of success, tried to adress these issues at the later stages of their projects, see Subsection~\ref{sec:interventions} (from Field notes -- observations, focus group interaction).

\subsubsection{Lack of communication}
We noted that a systematic lack of communication within the team  (e.g. \emph{standup meetings} and use of \emph{story boards}) plagued the groups throughout their respective projects. This added to the lack of understanding the merge conflicts as well as a lack of understanding the current project status. Further it added to \emph{waste} and \emph{loss of team productivity} when different pairs were working on the same task in parallel without knowing this. We also noted that the students became aware of these aspects and, to a varying degree of success, tried to address these issues at the later stages of their projects, see Subsection~\ref{sec:interventions}.  One group started using Trello instead of a physical story wall, while the others continued using story walls (from Field notes -- observations, focus group interaction).

\subsection{Consequences (F)}
This part of the analysis provides a reasoning on our observations of consequences of the phenomenon under study.
\subsubsection{Lack of understanding merge conflicts}
The systematic lack of understanding of merge conflicts surprised us, and it became the focus of the analysis. These merge conflicts obviously lead to a  loss of productivity, but it is not only limited to that. When going through the focus group material and the student reflections, we saw multiple examples of negatively loaded wording, indicating \emph{fear}, \emph{insecurity} and \emph{stress}. We find this to be clear indicators that issues with merge conflicts not only cause a loss of productivity in terms of linear time, but also that the absence of the needed tool support causes considerable cognitive load and stress on the developers.

\begin{itemize}
\item[S:] -- ``It is \emph{frightening} with a Wall of Text -- merge conflict/difference [indicating a very complicated merge] when in reality there is only a minor difference in a character or so [e.g. trailing space etc.]. In VS code you see both versions and you can simply choose what code [snippet] you want.''
\item[S:] -- ``You don't know how to revert changes in GIT you don't know if you will \emph{accidentally} [loss of control] replace/delete something [important]... you need to \emph{dare} to use GIT...''
\item[S:] -- ``\emph{uncertainty} results in many [of us] finding it \emph{stressful} with merge conflicts... when there is a "merge message" that just appears you don't really know what it means - will it result in overwrite - this makes it feel difficult, perhaps more so than it actually is...''
\end{itemize}(from Field notes -- saturation, questionnaires and focus group interaction).

\subsubsection{Absence of branch strategy and structure}
In addition to the systematic lack of understanding merge conflicts we also noted that branching itself was quite difficult for the teams. They had a hard time coming to grips with when to use separate branches (e.g. for bug fixes, tasks, stories and releases), when to close superfluous branches and branch naming conventions.

\begin{itemize}
\item[S:] -- ``It would have been better if we had used \emph{story specific branches}.''
\item[S:] -- ``We did not have a \emph{strategy for branching} from the beginning [of the project].''
\item[S:] -- ``We should have \emph{closed branches} that were no longer in use.''
\end{itemize}(from Field notes -- saturation, questionnaires and focus group interaction).

\subsubsection{Lack of project situation awareness}
Further we noted that there were issues in regards to understanding the current project situation/status. This included multiple pair working on the same tasks, different pair implementing similar utility functions, a lack of understanding of components in the projects, and ultimately not knowing whom to ask about implementation details.

\begin{itemize}
\item[S:] -- ``Lack of communication -- many of the problems we are facing would be solved if we would communicate better.''
\item[S:] -- ``People working on the same issue -- sometimes people work with solving the same problems without knowing it/each other.''
\item[S:] -- ``Lack of communication -- this lead to several interesting issues during sprint III where we went in different directions regarding architecture.''
\end{itemize}(from Field notes -- saturation, questionnaires and focus group interaction).

\subsection{Noted interventions (G)}
\label{sec:interventions}
We here describe the interventions implemented by the different teams as means to circumvent the issues they encountered in their projects. (from Field notes -- observations and focus group interaction).
\subsubsection{Creation of ``Git cheat sheet''}
We noted that the teams, after the first few sprints, realized that they needed a common manual for (and understanding of) basic Git operations. This was in most cases implemented as a \emph{spike} by a pair of team members in between sprints. Further, we saw this as an interesting example of knowledge transfer within the team.
\subsubsection{Improved documentation}
We noted that the all teams throughout the project started realising the importance of documentation. The observed interventions included a systematic way of describing commits (i.e. pointing out what story or what task had been worked on, rather than the initial, rather void, messages like \emph{'bugfix'}, \emph{'gui implementation'} etc.). We also noted that the teams started documenting the design of their architectures (using UML) and user interfaces (sketching on A3 paper). In addition we also noted that, while struggling with it in practice, all teams realised the importance of code documentation and made considerable attempts at documenting their code properly.
\subsubsection{Creation of basic process for branch/cm/releases}
We noted that all teams, after a few sprints started to develop a basic branch and configuration management process. This consisted of a more rigorous -- less ad hoc -- naming convention of branches, systematisation of main branch integration, and use of separate branches for stories, amongst other things. We do not consider the actual details as important as the observation that the teams, themselves, organically came to the conclusion that they needed a more systematic approach in regards to branching and configuration management. In addition we also noted that all teams, having experienced the value of explorative testing in the experiment presented in subsection~\ref{sec:experiment_testing}, started doing so well in advance of their releases.
\subsubsection{Improved communication}
We noted that all groups became aware of the need of improved communication. One team started using Trello as means of establishing a sound project overview. All teams further noticed the importance of standup meetings, and systematically started running more frequently.
\subsubsection{Experimentation with different IDEs}
As previously described we noted that two of the teams started exploring other IDEs in order to circumvent their perceived issues with Eclipse.

\section{Literature review}
\label{sec:literaturereview}
We will here briefly discuss our findings from extant literature in regards to Distributed Cognition, Git and Tool complexity.
\subsection{Distributed cognition}
Distributed cognition (DC) is a sub-discipline of studies of cognition in which the one of the traditional cornerstones of cognition -- ``that cognitive processes such as memory, decision making and  reasoning, are limited to the internal mental states of an individual''~\cite{hansen_dc} -- is questioned and rejected. Instead it argues that the social context of individuals as well as artefacts forms a cognitive system transcending the cognition of each individual involved~\cite{flor_dc}, i.e., a cognitive system extending beyond the mind of one single individual~\cite{mangalaraj_dc}.  The concept was pioneered by Hutchins who studied the cognitive activities on the navigation bridge of US naval vessels~\cite{hutchins_dc}.

Hollan, Hutchins and Kirsh extended DC into the realm of Human-Computer Interaction (HCI) as well as to some extent into Software Engineering, stating that a distributed cognitive process (or system) is ``delimited by the functional relations among the elements that are part of it, rather by the spatial colocation of the elements'', and that as a consequence ``at least three interesting kinds of distribution of cognitive processes become apparent: [a)] cognitive processes may be distributed across members of a social group[;] [b)] cognitive processes may involve coordination between internal and external (material or environmental) structure [and, c)] processes may be distributed through time in such a way that the products of earlier events can transform the nature of later events.''~\cite{hollan_dc}

Despite the fact that the theory of distributed cognition was suggested as a fruitful approach for investigating and explicating phenomenon related to software engineering several decades ago -- Flor and Hutchins empirically studied pair-programming from a distributed cognition perspective as early as 1991~\cite{flor_dc} -- few examples exist of actual software engineering studies using distributed cognition as scientific lens. In 2014, Mangalaraj et al.~\cite{mangalaraj_dc} highlighted Sharp and Robinson~\cite{sharp_distributed_1}, Hansen and Lyytinen~\cite{hansen_dc}, and Ramasubbu et al.~\cite{ramasubbu_dc} as ``the few notable exceptions'' of extant software engineering research utilising Distributed Cognition. To this list we would like to add Walenstein~\cite{walenstein_cognitive_diss}, a recent study by Buchan et al.~\cite{buchan_dc} and other works by Sharp et al.~\cite{sharp_ethnographic_xp}~\cite{sharp_role_dc}~\cite{sharp_role_2_dc}~\cite{sharp_distributed_2}~\cite{sharp_collaboration_dc}. 

\subsection{Git \& Merge}
Our literature findings in regards to user experience of Git were surprisingly limited. What we could find was three relevant papers: Church et al.~\cite{church_git}, Perez \& de Rosso~\cite{perez_de_rosso_git}, and de Rosso \& Jackson~\cite{de_rosso_jackson_git}.

We note that these papers, to some extent, validate our findings that Git is a very complex tool to use, and our conclusion is that there is considerable lacunae in literature in this regard. Future research should include a more thorough literature study in regards to Git and merge tools.

\subsection{Eclipse and tool complexity}
The issues related to tool complexity among novices are largely substantiated by extant literature -- Moody~\cite{moody_1} discussed the different levels of support needed by novices and experts when it comes to visual languages based on Cognitive Fit Theory~\cite{vessey_cft_1}. We can also see the same patterns in research on expertise by Chi et al.~\cite{chi_2}. Further, Storey et al.~\cite{storey_improving} as well as Rigby \& Thomson~\cite{rigby_thompson_1} have specifically described issues of novices in regards to Eclipse.

\section{Ethical considerations}
\label{sec:ethics}
With the study focus on groups rather than individual students, there was no legal need for formal ethical hearing under the jurisdiction under which this research was conducted. Only collection of \emph{sensitive} personal data, such as race or ethnic origin, political views, etc.\footnote{\url{https://www.staff.lu.se/research-and-education-0/research-support-0/research-ethics-and-animal-testing-ethics/ethical-review}} 
We did still submit and register a description of the study to the university's ethics review board. As the course is graded \emph{Pass} and \emph{Fail} only, and the only way students to fail is by considerable absence, we judged that there was no major issue with conflicting roles of researcher/teacher for the first author. In addition, our presence during sprints and planning sessions allowed the student groups more teacher time than what they would have experienced otherwise. Further, we stress again that all students were systematically offered to retire themselves from the groups being observed. 

Liebel \& Chakraborty present an updated  systematic mapping study on ethical issues in empirical software engineering studies using students~\cite{liebel_ethical_2021}, and highlight that study conditions and power relations between students and instructors are special areas of concern. We would like to stress that the consent to enter the study was informed, and we feel that we presented all students with the systematic ability to retire from the research, that we have been transparent with the conditions and that the power relations were, in reality, unaffected by the study condition since the role of the researchers were to act as customers in the actual projects.

The quasi-experiments we exposed the students to, had been used as improvised \emph{project disturbances} and as improvised exercises aimed at exploratory testing by the first author in previous years, and it appeared to make a sound addition to the learning outcome of the students.  Based on the fact that the learning outcome of the students was not compromised, that all data was collected with consent and anonymously, that the findings will benefit the students of the next instantiation of the course and the very high course evaluation grades the students awarded the course post completion, we do not feel that we have any ethical qualms in regards to the study. 

\section{Threats to validity}
The use of students as basis for research can be controversial \cite{students_1}\cite{students_2}\cite{students_3} from a generalisation perspective as well as from student privacy and learning perspectives. In terms of generalisation, H{\"o}st et al. highlight that students working under life-like circumstances serve can function as a reasonable proxy for real life settings/practitioners~\cite{students_exp}. In this study we selected students to capture a \emph{novice point of view}, thus providing us with a different perspective of causes of cognitive load drivers.  Further,  by acting as customers on site we were able to take a participatory observation position allowing us to some extent `blend', while retaining an analytical ethnographic stance~\cite{sharp_ethno}. 

In addition to discussing ethical dimensions of software engineering carried out on student populations, Liebel \& Chakraborty~\cite{liebel_ethical_2021} also discuss the scientific value of such studies. Highlighting that research conducted using qualitative methodologies such as \emph{case studies, observational studies and ethnography} the actual \emph{case context} is a  ``deciding factor'' and therefore cannot generally be separated from the ``studied phenomenon''. That being said, Stol \& Fitzgerald highlight the value of knowledge seeking research approaches such as ethnography~\cite{stol_abc_2018} using the work by Sharp \& Robinson~\cite{sharp_ethnographic_xp}  as an example of such research.

Somewhat tounge-in-cheek (and not drilling into the taxonomy of elephants, which is extensive and contextually important), we respond to the elephant/jungle metaphors provided by Stol \& Fitzgerald~\cite{stol_abc_2018} by stating that if you want to study juvenile elephants ridding themselves of bug infestation, and the methodologies deployed in such an activity you probably want to do it in the jungle~\footnote{Technically, we would, for (obvious) visibility reasons, prefer open plains rather than the jungle for observational studies. We will however not push the elephant metaphore further.}. If you, on the other hand, want to observe software development teams consisting of juniors solving software development issues, a computer hall at a university is, probably, about as ideal (and actually natural) as research environments come.
 
\begin{figure}[h]
  \centering
  \includegraphics[width=\linewidth]{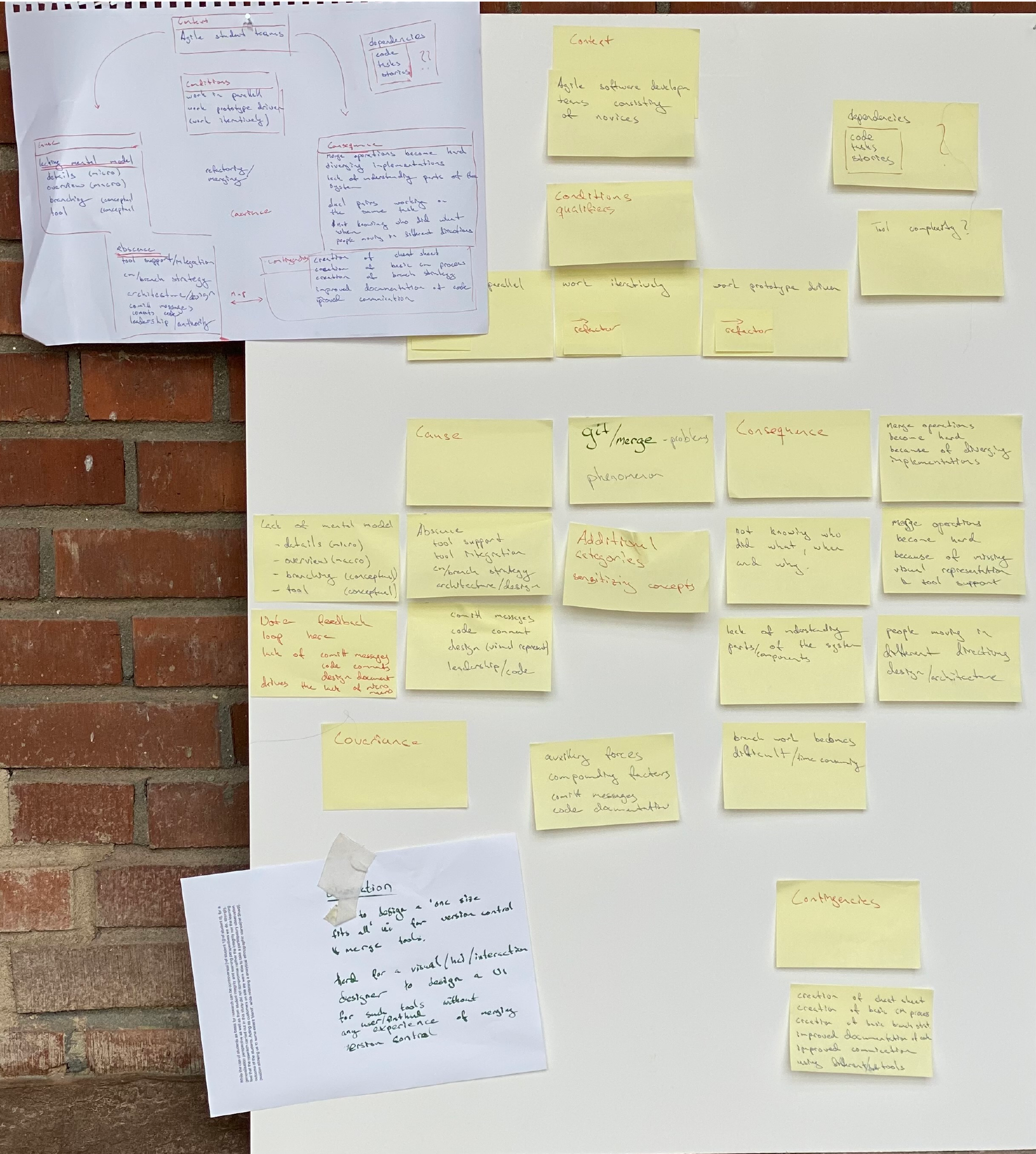}
  \caption{Post-it stickers used during focused coding}
  \label{fig:postit}
\end{figure}

\begin{figure}[h]
  \centering
  \includegraphics[width=\linewidth]{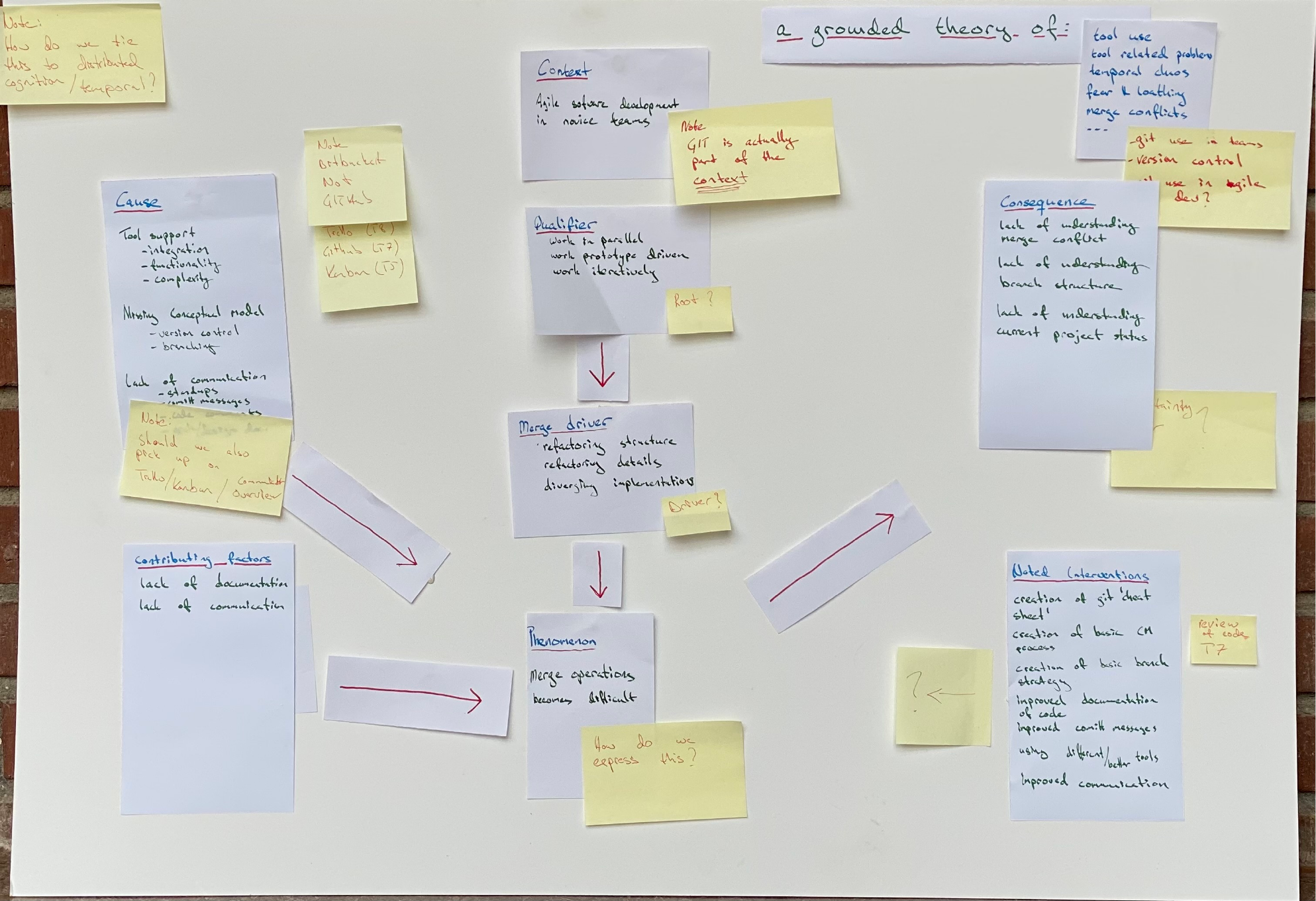}
  \caption{First rendering of our substansive grounded theory}
  \label{fig:firstgt}
\end{figure}

The procedures for the planning, data collection and analysis are reported in detail in Section~\ref{sec:Method}. In addition to the detailed procedural description, excerpts from our analysis artefacts used to generate our theory, are shown for transparency in Figures~\ref{fig:postit} \& \ref{fig:firstgt}. Figure~\ref{fig:postit} shows a cardboard sheet filled with post it stickers of incidents used during focused coding. Figure~\ref{fig:firstgt} shows a cardboard sheet with the first rendering of the theory.

GT studies are commonly evaluated based on the following criteria~\cite{charmaz_constructing}~\cite{stol_gt}:

\textbf{Credibility}: \emph{Is there enough data to merit claims of the study?} -- This study relies on the data set from one case study. The data set includes interviews, focus groups, observations and written reflections. The data set is quite extensive.

\textbf{Originality}: \emph{Do the perspectives offer new insight?} -- While cognitive load is not an unknown phenomenon in software engineering, we note that merge operations seem disproportianetly troublesome/difficult. We note a \emph{research gap} when it comes research on version control and merge operations.

\textbf{Usefulness}: \emph{Is the theory generated relevant for practitioners?} -- This study generates a theory that offers one explanation of how merge operations and branch work becomes difficult in projects. This can be used for reasoning on cognitive load in software engineering. The main contribution, in our opinion, is the observation of merge phenomenon and version control issues and the corresponding \emph{research gap}.

We take a pragmatic postpositivist~\cite{robson_real_2002} epistemological position in this paper. Our aim is to provide a grounded theory for reasoning on cognitive load in software engineering, using abductive reasoning on literature and data, and our ambition is to provide knowledge for software engineering research community and practitioners. We use grounded theory as a method, not an epistemological position. We acknowledge that all qualitative knowledge is inherently constructed. That being said, the phenomena we study do arguably exist, albeit in an artificial context largely unbound by natural laws. If the phenomena did not exist, there would be little point in studying them, nor their consequences on the human mind.

\section{Discussion and future work}
The findings in relation to the first research goal, \emph{to identify the most common cognitive load driver from the novice point of view}, was somewhat surprising. While we build our work on previous identification of the temporal perspective~\cite{helgesson_cogload}, the \emph{who, did what, when \& why}, we were quite surprised to see how large the impact of version control and merge operations were on the students. We also find it interesting to see the importance of tool support and functionality, tool integration and tool complexity in agile software development. To us the most interesting observation is the importance of visual merge support. We also noted that absence of communication and documentation was a contributing and confounding factor. We also note the absence of research on version control as an indicator for further research.

In addition to the codes described in our theory, we also noted other indications of cognitive load drivers in the material. The environment, in terms of ventilation on loud ambience were lamented on, describing the work situation as \emph{draining}. Further we also noted disruptions and task switching as a cause of concern -- described as a \emph{disruption of flow}.

We noted that distributed cognition, from our perspective, is indeed a sound lens for observing and analysing software development in agile teams and it is further interesting to note the reflections of \emph{history enriched objects} and \emph{temporal} cognitive dimensions made by Hollan et al.~\cite{hollan_dc}.

In regards to our second research goal, to chart what \emph{differences} or \emph{similarities} that can be observed between the different group compositions, we noted that there were  indeed observable, yet subtle differences between the different groups. With that said, during the field work we realised that the \emph{differences} we could observe, to us, were significantly less interesting than the \emph{similarities} we could observe. As a consequence we choose to use these similarities to strengthen the internal validity of our findings.

Following the 2020 iteration of the course during which these observations were made, the teachers working in the course had a few discussions on suitable interventions that could be extracted from the course. We added a more thorough introduction to Git and some harder actual hands on exercises as preparations for the 2021 course iteration. We further introduced more tool support to the students. While the Covid-19 situation has forced us to teach the course via Zoom and arguably made the whole course (that depends on teamwork) considerably more difficult we systematically noted that the students were suffering less from version control issues and were actually appearing to be more productive than previous years. For obvious reasons the pandemic situation prevented us from doing a more thorough follow up in the field. 

Further research in relation to this specific study will include a more focused study using direct observation of merge conflict resolution, and a 
benchmarking of a few popular IDEs in terms of version control integration and merge support, focusing on usability aspects. For a more long term perspective, \emph{mental models} and the temporal dimension of cognitive load in relation to software development tools could be used as a stepping stone for further research in terms of cognitive support for software development. The observed lacunae in extant literature in regards to version control, to us, is an indication that there is indeed relevant research to be done in this area, and the need of a systematic mapping study. It would further be interesting to do a critical analysis of research on literature on agile software development, possibly using meta-ethnography in conjunction with grounded theory.

\begin{acks}
We are thankful to the students who willingly contributed to the study through their coursework and by participating in focus groups. The work is conducted within the ELLIIT strategic research area (https://elliit.se).
\end{acks}

\bibliographystyle{ACM-Reference-Format}
\bibliography{Paper_2.bib}

\end{document}